\documentclass[prl,aps, superscriptaddress, twocolumn,showpacs,preprintnumbers,amsmath,amssymb, 10pt]{revtex4}
\usepackage{epsfig,amsmath}
\usepackage{amssymb,amsmath,wasysym}

\usepackage[english]{babel}
\usepackage{graphicx}
\begin{document}

\title{Spectral properties of fractional Fokker-Plank operator for the L\'evy flight in a harmonic potential}

\author{R. Toenjes}
\email{toenjes@uni-potsdam.de}
\affiliation{Institute of Physics and Astronomy, Potsdam University, 14476 Potsdam-Golm, Germany}
\author{I.M. Sokolov}
\affiliation{Institut f\"ur Physik, Humboldt Universit\"at zu Berlin, 12489 Berlin, Germany}
\author{E.B. Postnikov}
\affiliation{Department of Theoretical Physics, Kursk State University, 305000, Kursk, Russia}
\begin{abstract}
We present a detailed analysis of the eigenfunctions of the Fokker-Planck operator for the L\'evy-Ornstein-Uhlenbeck process, their asymptotic behavior and recurrence relations, explicit expressions in coordinate space for the special cases of the Ornstein-Uhlenbeck process with Gaussian and with Cauchy white noise and for the transformation kernel, which maps the fractional Fokker-Planck operator of the Cauchy-Ornstein-Uhlenbeck process to the non-fractional Fokker-Planck operator of the usual Gaussian Ornstein-Uhlenbeck process. We also describe how non-spectral relaxation can be observed in bounded random variables of the L\'evy-Ornstein-Uhlenbeck process and their correlation functions.
\end{abstract} 

\maketitle
\section{Introduction} 
This paper presents a detailed study of the recently reported non-trivial spectral properties of the fractional Fokker-Planck operator (FFPO) 
for the L\'evy-Ornstein-Uhlenbeck process (OUP) and the related unusual patterns of equilibration in such systems which were found
by the authors in \cite{Toenjes2013}. The Ornstein-Uhlenbeck process \cite{OrnsteinUhlenbeck1930} corresponding to the overdamped motion of a particle in a harmonic potential
under the influence of uncorrelated (white) noise is a paradigmatic model of non-equilibrium statistical physics describing relaxation and fluctuations
in a physical system in the vicinity of a stable equilibrium. In the original work Uhlenbeck and Ornstein considered the generic case of Gaussian noise. In recent years physicists have also focused on the more general case of systems under the influence of L\'evy noise and their description by fractional generalizations of the corresponding Fokker-Planck equations (FPE) \cite{Jespersen1999}. In this class of systems the Gaussian and the more  general L\'evy OUP are important, analytically tractable, stationary processes which are widely used for modeling, e.g. in neuroscience \cite{Lansky1995,SokolovLSG2006}, ecology \cite{ButlerKing2004,Worton1987}, finances and econophysics \cite{Barndorff2001,Linetsky2008} or in sensor systems \cite{Walton2004}. 
They have also been studied from a functional analysis point of view \cite{Peszat2011,Applebaum2004}, where their relation to certain problems 
of quantum field theory has been revealed. 
\\ \\
The relaxation of a dissipative system to its equilibrium state often shows a multiexponential pattern with
decay or relaxation rates which are typically considered to be independent of the initial condition and to follow
from the spectrum of the linear Fokker-Planck operator (FPO) describing the evolution of the probability density of the system. In one dimension the spectrum is obtained by a similarity transformation of the Fokker-Planck operator to a Hermitian operator \cite{Risken}. Since spectral methods are very useful for the description of relaxation processes, give deep theoretical
insights, and provide practical tools for numerical implementation of solution algorithms, understanding the properties of spectra and eigenfunctions of the process at hand, as well as pointing out possible pitfalls in using methods based on the spectral decomposition are of elemental importance. Ref. \cite{Toenjes2013} showed that the ``Hermitian'' spectral approach may break down even for the well-investigated case of the usual, Gaussian OUP. In particular, it was shown that modes with relaxation 
rates that are smaller than the smallest spectral relaxation rate may appear for initial conditions corresponding to 
probability densities which decay slowly as power laws at infinity. The presence of such modes is 
quite important since they can dominate the relaxation behavior over the whole time range. Even though for the L\'evy OUP there exists a similarity transformation of the fractional to the non-fractional FPO \cite{Toenjes2013} and, hence, also to the Hermitian quantum harmonic oscillator Hamiltonian, anharmonic ``non-spectral'' relaxation rates are typical for the general L\'evy OUP. The term ``non-spectral'' must not be understood in the sense that the corresponding rates do not belong to the spectrum 
of the FPO considered. On the contrary, as we proceed to show, they do. However, since the spaces of admissible functions
for the initial FPO and the corresponding Hermitian Schr\"odinger operator differ, the selection rules for
spectral components may differ as well, in spite of the existence of a similarity transformation mapping one operator onto the other one.
This fact stresses the difference between the situation pertinent to initial value problems for systems of linear homogeneous ordinary 
differential equations where the linear matrix operators connected by a similarity transformation are always isospectral, and for linear 
homogeneous partial differential equations like the FPE, where the transformation may map a function fully eligible
for the FPO onto one which does not belong to the domain of definition of the Schr\"odinger operator. 
\\ \\
In Ref. \cite{Toenjes2013} the authors considered the relaxation of the probability density function (PDF) to its equilibrium shape,
which can be done most conveniently in Fourier space. In the present work we concentrate on the real-space representations and discuss the asymptotic behavior of 
the eigenfunctions of the Fokker-Planck operators for Gaussian and L\'evy OUP which determines the spectrum of admissible eigenvalues. 
Furthermore we provide analytical expressions in real space for the eigenfunctions in case of the Gaussian OUP and the Cauchy OUP, 
and give the explicit form of the transformation kernel that maps the fractional Fokker-Planck operator of the Cauchy OUP to the non-fractional Fokker-Planck operator of the Gaussian OUP. 
In addition to these results, which are the direct continuation of \cite{Toenjes2013}, we discuss the relaxation 
behavior of properties other than the PDF, i.e. bounded observables which can be defined for \textit{any} initial condition, to their equilibrium values and their correlation functions at equilibrium. The fact that such observables and their correlation functions are experimentally accessible quantities makes the present work directly relevant for statistical data analysis. 
\section{L\'evy Ornstein Uhlenbeck Process}
\noindent
Let us first briefly recall the fractional Fokker-Planck equation for the L\'evy OUP \cite{Toenjes2013} and discuss the general properties of its solution.
Given a stochastic process that is subject to a linear restoring force and L\'evy white noise, the corresponding fractional Fokker Planck equation (FFPE) \cite{Jespersen1999} for the evolution of the probability density $p(x,t)$ reads
\begin{equation} \label{Eq:FFPE_natural_units}
	\partial_t p(x,t) = \nu\partial_x\left(xp\right) + D_\mu |\partial_x|^\mu p
\end{equation}
where the generalized restoring coefficient $\nu$ has the dimension of a frequency and the generalized diffusion constant $D_\mu$ has the dimension of a length to the fractional power of $\mu$ ($0<\mu\le 2$) per time. A linear rescaling to dimensionless units $\nu t\to t$ and $x(\nu D_\mu^{-1})^{1/\mu}\to x$ of time and space leaves the fraction $\mu$ of the derivate in the noise contribution as the only free parameter in the FFPE
\begin{equation}	\label{Eq:FFPEx}
	\partial_t p(x,t) = \partial_x\left(xp\right) + |\partial_x|^\mu p .
\end{equation}
The fractional derivative is defined as the symmetrized Riez-Weyl derivative by its action on the characteristic function $p(k,t)$ in Fourier-space where (\ref{Eq:FFPEx}) reads
\begin{equation} \label{Eq:FFPEk}
	\partial_t p(k,t) = -k\partial_k p - |k|^\mu p.
\end{equation}
Unless $\mu=2$ and $|\partial_x|^\mu=\partial_x^2$ is the ordinary second derivative, the fractional derivative, and thus the FFPO is only defined for functions that have a representation in Fourier space. As we will see, this restricts the eigenvalue spectrum of the FFPO to values $\lambda<1$, whereas the non fractional FPO has even and odd eigenfunctions for any $\lambda\in\mathbb{R}$. For the characteristic function $p^{st}(k)=\exp(-|k|^\mu/\mu)$ of the stationary probability density the the right hand side of (\ref{Eq:FFPEk}) evaluates to zero.
The exact time dependent solution of Eq. (\ref{Eq:FFPEk}) can be obtained via the method of characteristics for an arbitrary initial distribution with characteristic function $p(k,t)$  in the form
\begin{equation} \label{Eq:FFPEsolk}
	p(k,t+\tau) = p\left(ke^{-\tau},t\right) e^{-\frac{1}{\mu}|k|^\mu (1-e^{-\mu\tau})}.
\end{equation}
In \cite{Toenjes2013} we have studied the nonlinear rescaling of wave numbers $\kappa \sim \textrm{sign}(k)|k|^{\alpha}$, which is a linear transform of functions and generalized functions in Fourier space. Here we define
\begin{equation} \label{Eq:DefT}
	\left[T^\mu_\alpha p\right](\kappa,t) = \int_{-\infty}^\infty \delta\left(\alpha^\frac{1}{\mu}\textrm{sign}(\kappa)|\kappa|^{\frac{1}{\alpha}}-k\right) p(k,t) dk
\end{equation}
which transforms the fractional FPE of the L\'evy OUP changing the timescale and the order of the fractional derivative
\begin{equation} \label{Eq:FFPETrans}
	\frac{1}{\alpha} \partial_t \left[T^\mu_\alpha p\right](\kappa,t) = -\kappa \partial_\kappa \left[T^\mu_\alpha p\right] - |\kappa|^{\mu/\alpha} \left[T^\mu_\alpha p\right].
\end{equation}
For $\alpha=\mu/2$ we obtain the non-fractional FPE of the well known Gaussian OUP in Fourier representation
\begin{equation} \label{Eq:TransFPE}
	\frac{2}{\mu} \partial_t p(\kappa,t) = -\kappa \partial_\kappa p - \kappa^2 p.
\end{equation}
The harmonic or Hermitian spectrum of the FPO on the right hand side $\lambda_n=-n$ follows from similarity to the Hermitian quantum harmonic oscillator Hamiltonian (see next section). The eigenvalues in the harmonic spectrum of the original fractional FPO (Eq.~(\ref{Eq:FFPEk})) are therefore $\lambda_n = -n\mu/2$. 
\\ \\
The transformation (\ref{Eq:DefT}) affects the scaling of the characteristic functions in the vicinity of $k=0$ and thus integrability and asymptotic behavior of the corresponding functions in real space as $|x|\to\infty$.
While the value of $f(k=0)=[T^\mu_\alpha f](\kappa=0)$ is unaffected by the transformation, the corresponding integrals in real space $\int_{-\infty}^\infty f(x)dx=\int_{-\infty}^\infty [T^\mu_\alpha f](\chi)d\chi$ may only be defined by principal value. The inverse transformation of $T^\mu_\alpha$ is $T^{\mu/\alpha}_{1/\alpha}$. In \cite{Toenjes2013} we had defined $T_\alpha=\delta(\textrm{sign}(k)|k|^{\alpha}-k)$ which here corresponds to $T^\infty_\alpha$ and has $T^\infty_{1/\alpha}$ as its inverse.
The eigenvalue problem of the fractional FPO in Fourier space
\begin{equation} \label{Eq:FPOEigenvalueProblem}
	-k\partial_k \varphi_\lambda - |k|^\mu \varphi_\lambda = \lambda \varphi_\lambda
\end{equation}
is solved by a linear combination of even and odd eigenfunctions
\begin{equation} \label{Eq:EvenEigenfunc}
	\varphi_\lambda^+ = |k|^{-\lambda}e^{-\frac{1}{\mu}|k|^\mu}
\end{equation}
and
\begin{equation} \label{Eq:OddEigenfunc}
	\varphi_\lambda^- = i\textrm{sign}(k)|k|^{-\lambda}e^{-\frac{1}{\mu}|k|^\mu}.
\end{equation}
Eigenfunctions with $\lambda<0$ correspond to modes of the PDF decaying exponentially in time. The even eigenfunction $\varphi_{\lambda=0}^+(k)$ is the characteristic function of the unique staionary PDF solution of the FFPE (\ref{Eq:FFPEx}). The value and derivative of $\varphi^-_{\lambda=0}(k=0)$ and of $\varphi^\pm_{\lambda>0}(k=0)$ at $k=0$ are not well defined. This means that the improper integral of the eigenfunctions from negative to positive infinity does not exist in real space, i.e. they are no longer integrable there. For values $\lambda<1/2$, however, the eigenfunctions are still {\em square integrable} in Fourier space, and therefore also in real space. In that case they decay slower to zero at infinity than $1/x$  but faster than $1/\sqrt{x}$.
For $\lambda<1$ the eigenfunctions are {\em integrable} in Fourier space and the Riemann-Lebesgue Lemma provides $\varphi^\pm_{\lambda<1}(x)\to 0$ as $x\to\infty$. For $\lambda\ge 1$ the eigenfunctions (\ref{Eq:EvenEigenfunc},\ref{Eq:OddEigenfunc}) in Fourier space are no longer integrable and have no counterpart in real space. Eigenfunctions in real space with eigenvalues $\lambda\ge 1$ can be found in the non-fractional case with $\mu=2$. They correspond to generalized functions in Fourier space that are no longer given by (\ref{Eq:EvenEigenfunc},\ref{Eq:OddEigenfunc}). For other values $0<\mu<2$ a continuation to eigenvalues $\lambda\ge 1$ would require an alternative definition of the fractional derivative without refering to its action in Fourier space, possibly based on recurrence relations between eigenfunctions. We are not discussing this approach here. Which of the abundance of eigenvalues and eigenfunctions constitute a complete set depends on the physical problem at hand, not on similarity to a Hermitian operator. In contrast to Hermitian operator theory for square integrable functions with well defined inner product, there is no bi-orthogonal set of eigenfunctions of the FFPO and its formal adjoint, the usual analytic tool of spectral decomposition. Moreover, exchangability of spectral series and convolution, e.g. of the Green's function, is not guaranteed.
\\ \\
From their Fourier representation we can derive recurrence relations for the eigenfunctions with $\lambda<1$.
Any symmetrized fractional Riez-Weil derivative of an eigenfunction is again an eigenfunction of the FFPO
\begin{equation} \label{Eq:ShiftDownAlpha}
	|\partial_x|^{\alpha} \varphi^\pm_\lambda = -\varphi^\pm_{\lambda-\alpha}.
\end{equation}
In particular the symmetrized fractional derivative of order $\mu$ can be expressed rearranging (\ref{Eq:FFPEx}) for an eigenfunction as
\begin{equation} \label{Eq:ShiftDownMu}
	|\partial_x|^{\mu} \varphi^\pm_\lambda = -\varphi^\pm_{\lambda-\mu} = \left[(\lambda-1) - x\partial_x\right] \varphi^\pm_\lambda.
\end{equation}
The usual non-fractional and non-symmetrized derivative changes both the parity and the eigenvalue of an eigenfunction
\begin{equation} \label{Eq:FlipShift}
	\partial_x \varphi^\pm_\lambda = \mp\varphi^\mp_{\lambda-1}.
\end{equation}
Equations (\ref{Eq:ShiftDownAlpha},\ref{Eq:FlipShift}) can be combined to obtain the general recurrence relation between the eigenfunctions of the FFPO in real space
\begin{equation}	\label{Eq:Recurrence_x}
	\varphi^\pm_{\lambda-\mu} = (1-\lambda)\varphi_\lambda^\pm \mp x \varphi^\mp_{\lambda-1}
\end{equation}
and in Fourier space
\begin{equation}	\label{Eq:Recurrence_k}
	\varphi^\pm_{\lambda-\mu} = (1-\lambda)\varphi_\lambda^\pm \pm i\frac{d}{dk} \varphi^\mp_{\lambda-1}
\end{equation}
The eigenvalues $\lambda_{mn} = -(m+\mu n)$ with $m,n\ge 0$ are special since the conditional probability density $p(x t+\tau|x(t)=x_0)$ of the L\'evy OUP, which is the Green's function for the FFPE (\ref{Eq:FFPEx}) has an absolutely convergent expansion into the corresponding eigenfunctions \cite{Toenjes2013}. In this expansion the initial coordinate $x_0$ appears in powers of $m$ so that the convolution with a L\'evy stable initial distribution is not defined for all terms in the expansion. 
Equations (\ref{Eq:ShiftDownMu},\ref{Eq:FlipShift}) describe a constructive method to generate the corresponding eigenfunctions $\varphi^\pm_{mn}$ from the two stationary functions $\varphi^\pm_{00}$ as
\begin{equation} \label{Eq:Recurrence_A}
	\varphi^\pm_{m n+1} = \left[(m+1+\mu n) + x\partial_x\right] \varphi^\pm_{mn}
\end{equation}
and
\begin{equation} \label{Eq:Recurrence_B}
	\varphi^\pm_{m+1 n} = \pm \partial_x \varphi^\mp_{mn},
\end{equation}
i.e. by multiplication with $x$ and taking simple derivatives. 
\\ \\
The large $x$ asymptotics of functions in real space follows from their behavior at small wave numbers $k$ in Fourier space. Only if the $n$th derivative of a function $f(k)$ with respect to $k$ at zero exists, the proper integral of $x^nf(x)$ over the whole real line exists, as well. If this integral is finite for all $n$, the function $f(x)$ must decay faster at infinity than any power law. Conversely, if the function decays to zero at infinity as a power law or slower it is not smooth at $k=0$ in Fourier space. Using (\ref{Eq:Recurrence_k}) the derivative of an eigenfunction of the FFPO with respect to $k$ can be expressed as
\begin{equation} \label{Eq:PartialK}
	\frac{d}{dk}\varphi^\pm_\lambda = \pm i\left(\lambda\varphi^\mp_{\lambda+1} + \varphi^\mp_{\lambda+(1-\mu)}\right).
\end{equation}
and in particular
\begin{equation} \label{Eq:PartialK0}
	\frac{d}{dk}\varphi^\pm_0 = \pm i\varphi^\mp_{1-\mu}.
\end{equation}
For negative non-integer values of $\mu$ or $\lambda$  repeated differentiation will eventually generate an eigenfunction corresponding to a positive eigenvalue, for which $\varphi^\pm_{\lambda>0}(k=0)$ is not defined. For non positive, integer values $\lambda$ and $\mu=2$, i.e. for the usual Gaussian OUP, differentiation switches the parity of the eigenfunction and changes the eigenvalue from even to odd and vice versa. The space of even eigenfunctions with $\lambda=-2n$ and odd eigenfunctions with $\lambda=-(2n+1)$ and $n\in\mathbb{N}$ is therefore invariant under differentiation with respect to $k$. Only these eigenfunctions are well defined and have derivatives of all orders at $k=0$. Derivatives of even eigenfunctions with odd eigenvalues and odd eigenfunctions with even eigenvalues will eventually have a component $\varphi^-_0$, which is not well defined at $k=0$ and for which higher derivatives do not exist. In case of the Cauchy OUP with $\mu=1$ and for non positive, integer values $\lambda$, repeated derivation will also eventually have a component $\varphi^-_0$. In conclusion, the eigenfunctions of the FFPO of the L\'evy OUP with $\lambda<1$ in general decay as power laws at infinity except the eigenfunctions of the FPO for the Gaussian OUP corresponding to its  harmonic or Hermitian spectrum \cite{Toenjes2013}. In real space the eigenfunctions can be expressed in terms of Fox H-functions \cite{Jespersen1999} which reduce to simpler functions for special values of $\mu$ and $\lambda$, as will be discussed in the following sections.
\section{Quantum Harmonic Oscillator and OU Process}
In this section we present two derivations for the spectrum and the eigenfunctions of the quantum harmonic oscillator, based on the transformation of the Schroedinger equation to a Fokker-Planck equation and an asymptotic analysis of the more easily obtained eigenfunctions of the FPO. Finding admissible eigenfunctions from a scaling at small wave numbers in Fourier space and bringing these eigenfunctions into the form of  Rodrigues' Formula for Herminte polynomials with Gaussian envelope is a particularly instructional alternative to the usual and more involved derivation via Sommerfeld Series expansion \cite{Sommerfeld}. A transformation of the FPE into Kummer's hypergeometric differential equation, on the other hand \cite{KuznetsovThes}, gives direct, analytic real space representations of the eigenfunctions of the FPO in terms of Kummer and Tricomi functions with known asymptotics and reductions to Hermite polynomials in the case of admissible eigenfunctions and eigenvalues of the harmonic spectrum. The method of transformation of the stationary Schr\"odinger equation to a partial differential equation that can readily be solved in Fourier space has been applied in \cite{PalmerRaff2011} for a broader class of quantum systems. However, there the authors avoided a detailed discussion of the solution asymptotics arguing that the selection rule for bound states is equivalent to the requirement of single valuedness of quantum mechanical wave functions \cite{Merzbacher1962}. 
\\ \\
Let us consider the dimensionless, stationary Schroedinger equation
\begin{equation} \label{Eq:SEq}
	\mathrm{H}\psi = \left[V(x) - \frac{\partial^2}{\partial x^2}\right] \psi = \epsilon \psi 
\end{equation}
or equivalently
\begin{equation} \label{Eq:SeqOrder2ODE}
	\frac{\partial^2}{\partial x^2} \psi = \left[V(x)-\epsilon\right] \psi
\end{equation}
with Hamiltonian operator $\mathrm{H}$, one dimensional potential $V_{\min}\le V(x)<\infty$, diverging at $\pm\infty$ and eigenvalues $\epsilon$. 
The argument, that the spectrum of this Hamiltonian is discrete and bounded from below, i.e. $\epsilon=\epsilon_n>\epsilon_0$, is usually given as follows.
For any given $\epsilon$ there is a two dimensional linear space of fundamental solutions of this homogeneous, second order linear differential equation. Since $\epsilon<V(x)$ for large enough $x$, the fundamental solutions of (\ref{Eq:SeqOrder2ODE}) are repelled by the $x$-axis and in general grow to $\pm\infty$. For $\epsilon\le V_{\min}$ only the zero solution $\psi=0$ does not diverge at infinity. For $\epsilon>V_{\min}$ there is a region where the solutions bend towards the $x$-axis and there exists exactly one subspace where the solution goes to zero at infinity and another subspace where the solution goes to zero at negative infinity. These two subspaces are in general not identical, except for discrete values $\epsilon_n$. The ground state solution $\psi_0$, corresponding to the smallest value $\epsilon_0>V_{\min}$ can be chosen strictly positive. 
We will now consider the transformation
\begin{equation} \label{Eq:SEqFPETrans}
	\varphi = \psi_0 \psi
\end{equation}
changing the stationary Schroedinger Equation (\ref{Eq:SEq}) into the eigenvalue problem for a Fokker-Planck operator $\mathrm{L}$
\begin{equation}	\label{Eq:1dFPOEigenvalueProblem}
	\mathrm{L}\varphi_\lambda = \frac{\partial}{\partial x}\left(U'\varphi_\lambda\right) + \frac{\partial^2}{\partial x^2} \varphi_\lambda = \lambda \varphi_\lambda.
\end{equation}
with $\lambda=\epsilon_0-\epsilon$ for the diffusion in a potential
\begin{equation} \label{Eq:DefU}
	U = -2\log\psi_0.
\end{equation}
This is the Cole-Hopf transformation that changes the nonlinear differential equation
\begin{equation} \label{Eq:NlDEq}
	V-\epsilon_0 = \frac{1}{2}U'^2-\frac{1}{4}U''
\end{equation}
into the Schroedinger Equation (\ref{Eq:SEq}) for the ground state. Every eigenfunction $\psi_n$ of the stationary Schroedinger equation with eigenvalue $\epsilon_n$ and $\psi_n(\pm\infty)=0$ has a corresponding eigenfunction $p_n=\varphi_{\lambda_n}$ of the Fokker-Planck operator, decaying to zero at infinity, with eigenvalue $\lambda_n=(\epsilon_0-\epsilon_n)$. However, the converse is not necessarily true. There can be eigenfunctions of the Fokker-Plank operator which decay to zero at infinity and correspond to eigenfunctions $\psi_\epsilon$ of the Hamiltonian that diverge at infinity. The space of solutions of the stationary Schroedinger equation which decay to zero at infinity is therefore homomorph to only a subspace of the solutions of the eigenvalue problem of the corresponding Fokker-Planck operator under the same boundary conditions. We will see this explicitly in the case of a quadratic potential, i.e. the Hamiltonian of the quantum harmonic oscillator and the Fokker-Planck operator of the classical Ornstein Uhlenbeck process. Here we have, after an appropriate scaling
\begin{equation} \label{Eq:harmonicU}
	U=\frac{1}{2} x^2
\end{equation}
and
\begin{equation} \label{Eq:QMPotential}
	V-\epsilon_0 = \frac{1}{2}U'^2 -\frac{1}{4}U'' = \frac{1}{2} \left(x^2 - 1\right).
\end{equation}
Since $\psi_0=\exp(-U/2)=\exp(-x^2/4)$ is by construction the ground state of the stationary Schroedinger Equation with potential $V(x)$, the transformation (\ref{Eq:SEqFPETrans}) leads via (\ref{Eq:1dFPOEigenvalueProblem}) to the eigenvalue problem of the non fractional FPO of the Gaussian OUP
\begin{equation} \label{Eq:GaussFPOEigenvalueProblem}
	\frac{\partial}{\partial x}\left(x\varphi_\lambda\right) + \frac{\partial^2}{\partial x^2} \varphi_\lambda = \lambda \varphi_\lambda.
\end{equation}
Following the asymptotic analysis in the first section, we find that with $n\in\mathbb{N}$ only even eigenfunctions 
\begin{equation} \label{Eq:EvenPk}
	p_{2n}(k) = \varphi^+_{-2n} = k^{2n} \exp\left(-\frac{1}{2}k^2\right)
\end{equation}
corresponding to even non positive eigenvalues and odd eigenfunctions 
\begin{equation} \label{Eq:OddPk}
	p_{2n+1}(k) = \varphi^-_{-(2n+1)} = ik^{2n+1} \exp\left(-\frac{1}{2}k^2\right)
\end{equation}
corresponding to odd negative eigenvalues decay faster at infinity than any power law. In all other cases $\psi_\epsilon = \varphi_\lambda/\psi_0 = \varphi_{\lambda} \exp(x^2/4)$ diverges. 
Multiplication in Fourier space in (\ref{Eq:EvenPk},\ref{Eq:OddPk}) with $(-1)^nk^{2n}$ and $(-1)^{n+1}ik^{2n+1}$ respectively correspond to derivatives in real space.
Thus the eigenfunctions $\psi_n = p_n / \psi_0$ of the quantum harmonic oscillator Hamiltonian follow directly from the Rodrigues' Formula for Hermite Polynomials in the ``probabalistic'' notation \cite{Abramowitz}
\begin{equation}
	\textnormal{\em He}_n(x) = (-1)^n e^{\frac{1}{2}x^2}\frac{d^n}{dx^n}e^{-\frac{1}{2}x^2}
\end{equation}
as
\begin{eqnarray}	
	\psi_{2n} &=& \frac{1}{\sqrt{2\pi}}e^{\frac{1}{4}x^2}(-1)^n \frac{d^{2n}}{dx^{2n}} e^{-\frac{1}{2}x^2} \label{Eq:RodriguesEven} \\ \nonumber \\
		  &=&  \frac{(-1)^n}{\sqrt{2\pi}}  e^{-\frac{1}{4}x^2} \textnormal{\em He}_{2n}(x), \nonumber \\ \nonumber \\
	\psi_{2n+1} &=& \frac{1}{\sqrt{2\pi}}e^{\frac{1}{4}x^2}(-1)^{n+1} \frac{d^{2n+1}}{dx^{2n+1}} e^{-\frac{1}{2}x^2} \label{Eq:RodriguesOdd} \\ \nonumber \\
		&=&  \frac{(-1)^n}{\sqrt{2\pi}}  e^{-\frac{1}{4}x^2} \textnormal{\em He}_{2n+1}(x), \nonumber
\end{eqnarray}
These are of course the well known eigenstates consisting of Hermite polynomials with Gaussian envelope, decaying to zero at infinity as required.
\begin{figure}[t!] 
\includegraphics[width=4.2cm]{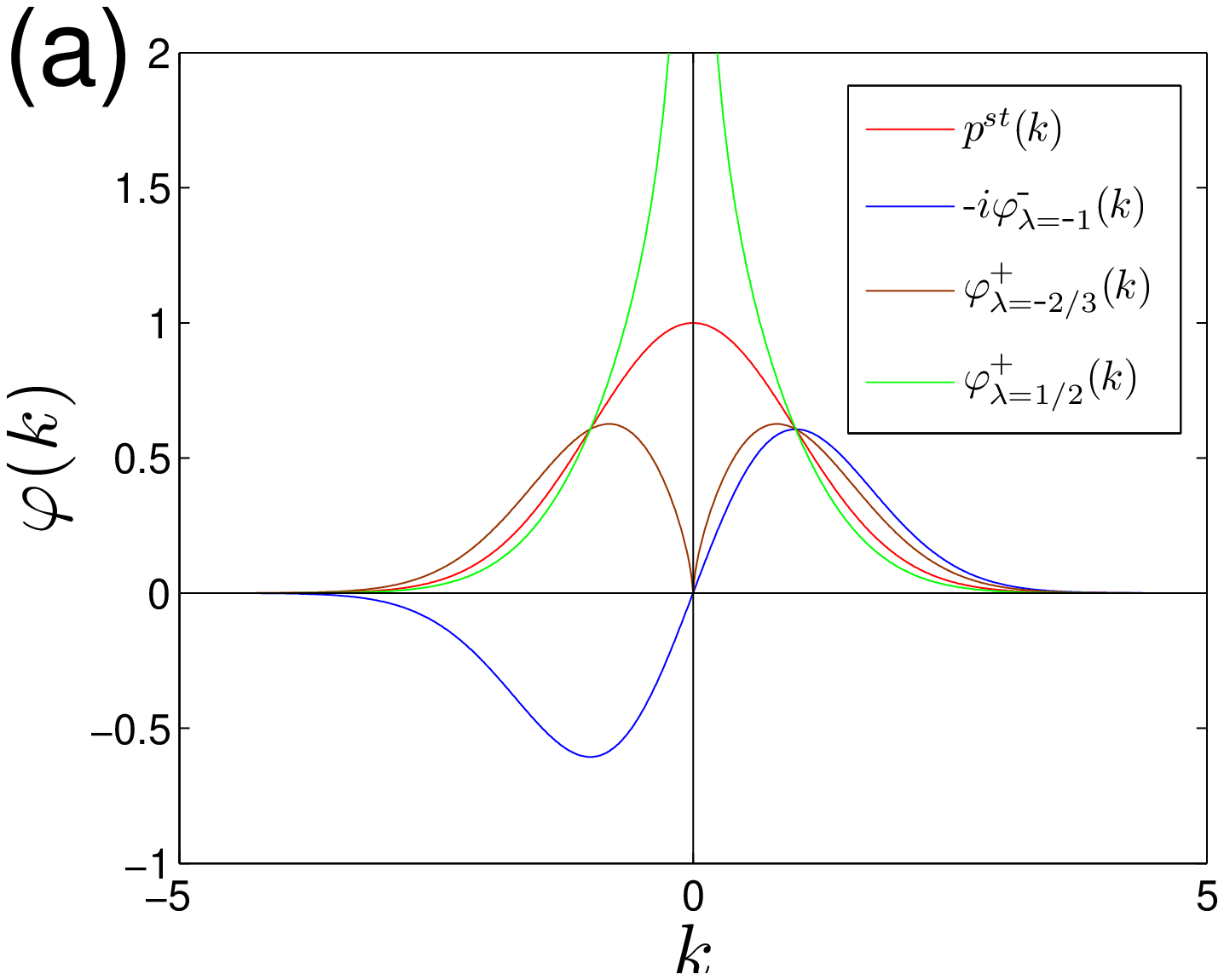}
\includegraphics[width=4.2cm]{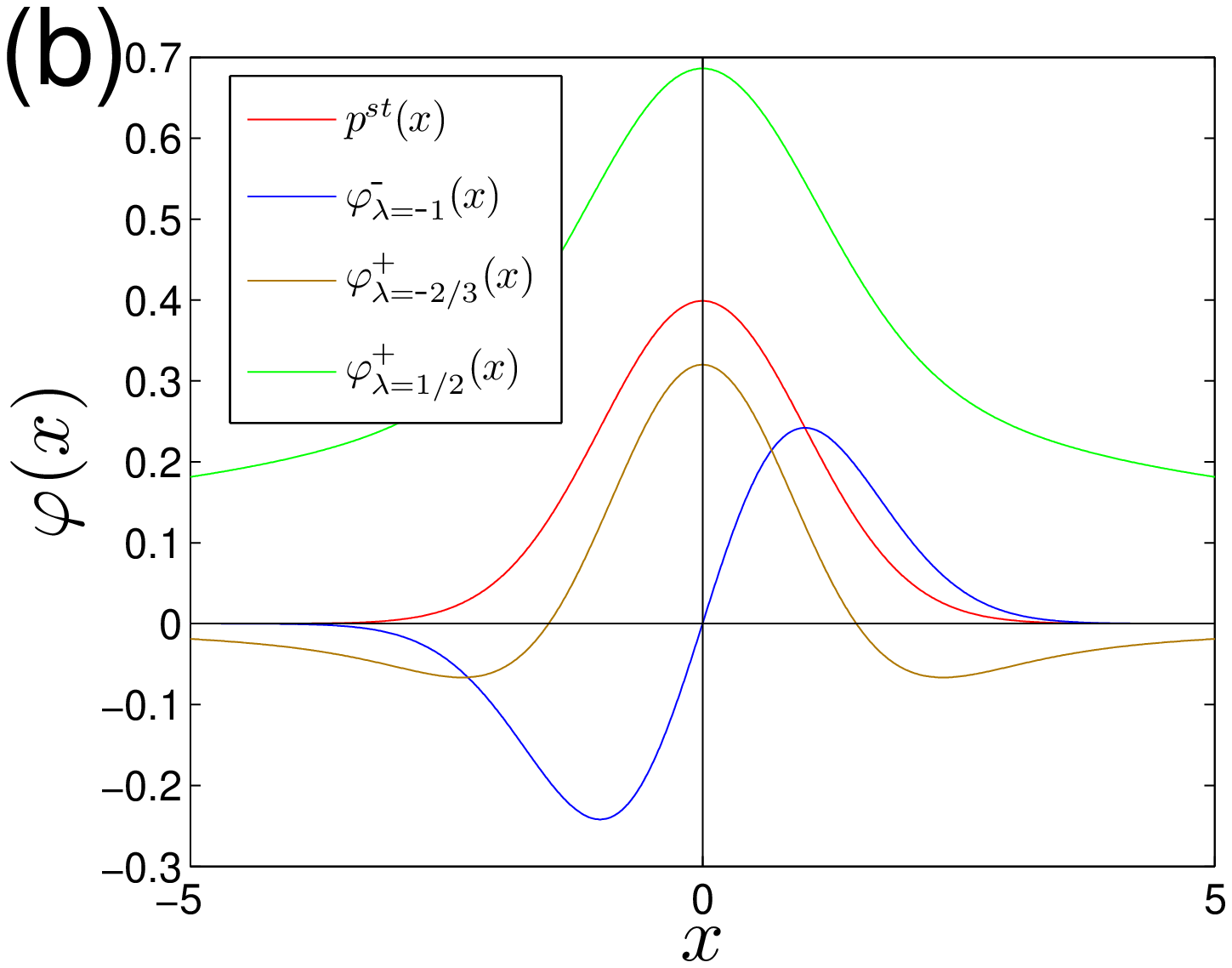}
\caption{Eigenfunctions (\ref{Eq:GaussFPOEigenfuncAppendix}) of the Gaussian OUP FP operator (a) in Fourier space and (b) in real space. Shown are the eigenfunctions corresponding to the eigenvalues $\lambda=0$ (even, stationary distribution, red), the first odd harmonic eigenfunction with $\lambda=-1$ (blue), the even eigenfunction with $\lambda=-2/3$ (brown) and the even eigenfunction with $\lambda=1/2$ (green). Only $p^{st}(x)$ and $\varphi^-_{\lambda=-1}$ are localized with a gaussian envelope in real space. All other eigenfunctions decay asymptotically as power laws.}
\label{Fig:GaussEigen}
\end{figure}
\\ \\
A more formal argument gives the real space representation of all eigenfunctions of the FPO for the Gaussian OUP for any eigenvalue $\lambda\in\mathbb{R}$, in terms of Kummer and Tricomi confluent hypergeometric functions.
Substituting $z=-x^2/2$ and $w(z)=\varphi(\sqrt{-2z})$ turns Eq. (\ref{Eq:GaussFPOEigenvalueProblem}) into Kummer's hypergeometric differential equation \cite{KuznetsovThes}
\begin{equation} \label{Eq:KummerHDEq}
z\frac{d^2w}{dz^2}+\left[\frac{1}{2}-z\right]\frac{dw}{dz}-\frac{1}{2}\left[1-\lambda\right]w=0,
\end{equation}
with a general solution \cite{Abramowitz} that is a linear combination of 
\begin{equation} \label{Eq:KummerFunc}
M\left(\frac{1}{2}[1-\lambda],\frac{1}{2},z\right)=M\left(\frac{1}{2}\left[1-\lambda\right],\frac{1}{2},-\frac{x^2}{2}\right)
\end{equation}
and
\begin{eqnarray} \label{Eq:TricomiFunc}
U\left(\frac{1}{2}[1-\lambda],\frac{1}{2},z\right)&=&\sqrt{\pi}\left\{\frac{M\left(\frac{1}{2}\left[1-\lambda\right],\frac{1}{2},-\frac{x^2}{2}\right)}{\Gamma\left(1-\frac{\lambda}{2}\right)}\right. \\
&&-\left.ix\sqrt{2}\frac{M\left(1-\frac{\lambda}{2},\frac{3}{2},-\frac{x^2}{2}\right)}{\Gamma\left(\frac{1}{2}\left[1-\lambda\right]\right)}\right\}, \nonumber
\end{eqnarray}
where $M(a,b,z)$ and $U(a,b,z)$ are Kummer and the Tricomi functions. We can select real-valued symmetric and antisymmetric components to represent the even and odd eigenfunctions of the original FPO
\begin{equation} \label{Eq:EvenPhiGauss}
\varphi^+_{\lambda}(x)\sim M\left(\frac{1}{2}\left[1-\lambda\right],\frac{1}{2},-\frac{x^2}{2}\right)
\end{equation}
and
\begin{equation} \label{Eq:OddPhiGauss}
\varphi^-_{\lambda}(x)\sim xM\left(1-\frac{\lambda}{2},\frac{3}{2},-\frac{x^2}{2}\right).
\end{equation}
Applying the Kummer transformation $M(a,b,z)=M(b-a,b,-z)\exp(z)$ \cite{Abramowitz}, for even and odd non positive integer values of $\lambda$ the eigenfunctions reduce to the special cases 
\begin{eqnarray} \label{Eq:EvenPx}
p^+_{2n}(x)&\sim&e^{-\frac{x^2}{2}}M\left(-n,\frac{1}{2},\frac{x^2}{2}\right) \nonumber \\ \\
		&&=\frac{(-2)^nn!}{(2n)!}e^{-\frac{x^2}{2}}\textnormal{\em He}_{2n}(x) \nonumber
\end{eqnarray}
and
\begin{eqnarray} \label{Eq:OddPx}
p^-_{2n+1}(x)&\sim&xe^{-\frac{x^2}{2}}M\left(-n,\frac{3}{2},\frac{x^2}{2}\right) \nonumber \\ \\
		&&=\frac{(-2)^nn!}{(2n+1)!}e^{-\frac{x^2}{2}}\textnormal{\em He}_{2n+1}(x). \nonumber
\end{eqnarray}
Again, one finds, the eigenfunctions $\psi_n = p_n \exp(x^2/4)$ are the Hermite polynomials modulated by a Gaussian envelope $\exp(-x^2/4)$.
On the other hand, for fractional values of $\lambda$ the eigenfunctions (\ref{Eq:EvenPhiGauss},\ref{Eq:OddPhiGauss}) scale at infinity \cite{Abramowitz} as
\begin{equation}\label{Eq:EigenAsymptotics}
\varphi^\pm_\lambda(x) \sim |x|^{\lambda-1}.
\end{equation}
The corresponding eigenfunctions $\psi_{\lambda}(x)\sim |x|^{\lambda-1}\exp(x^2/4)$ of the quantum harmonic oscillator Hamiltonian diverge. Non integer values are therefore not part of the harmonic or Hermitian spectrum of the FPO for the Gaussian OUP. Even though the inverse Fourier transform of (\ref{Eq:EvenEigenfunc},\ref{Eq:OddEigenfunc}) only exists for $\lambda<1$, the eigenfunctions (\ref{Eq:EvenPhiGauss}) and (\ref{Eq:OddPhiGauss}) are well defined for any real $\lambda\in\mathbb{R}$. For $\lambda=1$ the eigenvalue problem of the FPO is solved by a linear combination of
\begin{equation}
	\varphi^+_1 = 1
\end{equation}
and the error function
\begin{equation}
	\varphi^-_1 = \sqrt{\frac{\pi}{2}}\textrm{erf}\left(\frac{x}{\sqrt{2}}\right).
\end{equation}
The even and odd eigenfunctions for respectively odd and even positive integer eigenvalues are given by generalized Laguerre polynomials \cite{Abramowitz}
\begin{eqnarray}
	\varphi^+_{2n+1}(x) &\sim& M\left(-n,\frac{1}{2},-\frac{1}{2}x^2\right) \nonumber \\ \\
				&&	= \frac{n!}{\left(\frac{1}{2}\right)_n} L^{-1/2}_n\left(-\frac{1}{2}x^2\right) \nonumber
\end{eqnarray}
\begin{eqnarray}
	\varphi^-_{2n+2}(x) &\sim& xM\left(-n,\frac{3}{2},-\frac{1}{2}x^2\right) \nonumber \\ \\
				&&	= \frac{n!}{\left(\frac{3}{2}\right)_n} x L^{1/2}_n\left(-\frac{1}{2}x^2\right) \nonumber
\end{eqnarray}
where $(a)_n=\Gamma(a+n)/\Gamma(a)$ is the Pochhammer symbol. 
\section{Cauchy Ornstein Uhlenbeck process}
In the special case of $\mu=1$ the integrated white L\'evy noise is a Cauchy process. Subjected to an additional linear restoring force one observes a Cauchy Ornstein-Uhlenbeck process. 
Using Euler's Integral identity for the Gamma function and $\lambda<1$ \cite{Abramowitz} we have
\begin{eqnarray} \label{Eq:COU_LaplTrans}
	\frac{1}{2\pi}\int_{0}^\infty k^{-\lambda} e^{-k-ikx} dk 	&=& \frac{1}{2\pi}\left(1+ix\right)^{\lambda-1} \Gamma\left(1-\lambda\right) \nonumber \\ \\
								&=& \frac{1}{2\pi}\left(\frac{1-i x}{x^2+1}\right)^{1-\lambda}\Gamma\left(1-\lambda\right). \nonumber
\end{eqnarray}
The Fourier transforms of $\varphi_\lambda^\pm(k)$ (Eqs.\ref{Eq:EvenEigenfunc},\ref{Eq:OddEigenfunc}) are equal to respectively twice the real and the negative imaginary part of that integral, i.e.
\begin{eqnarray} \label{Eq:COUPEigen}
	\varphi_\lambda^+(x) 	&=& \frac{1}{\pi}\Gamma\left(1-\lambda\right)\left(\frac{1}{x^2+1}\right)^{\frac{1-\lambda}{2}}\cos\left((1-\lambda)\arctan x\right), \nonumber \\ \\
	\varphi_\lambda^-(x)    &=& \frac{1}{\pi}\Gamma\left(1-\lambda\right)\left(\frac{1}{x^2+1}\right)^{\frac{1-\lambda}{2}}\sin\left((1-\lambda)\arctan x\right). \nonumber
\end{eqnarray}
They assume the form of rational functions for non-positive integer eigenvalues
\begin{eqnarray} \label{Eq:COUPEigenInteger}
	\varphi_{-n}^+(x) 	&=& \frac{n!}{\pi} \frac{\textrm{Re}\left[(1-ix)^{n+1}\right]}{(x^2+1)^{n+1}} \nonumber \\ \nonumber \\
	\varphi_{-n}^+(x) 	&=& -\frac{n!}{\pi} \frac{\textrm{Im}\left[(1-ix)^{n+1}\right]}{(x^2+1)^{n+1}}. \nonumber
\end{eqnarray}
\begin{figure}[t!] 
\includegraphics[width=4.2cm]{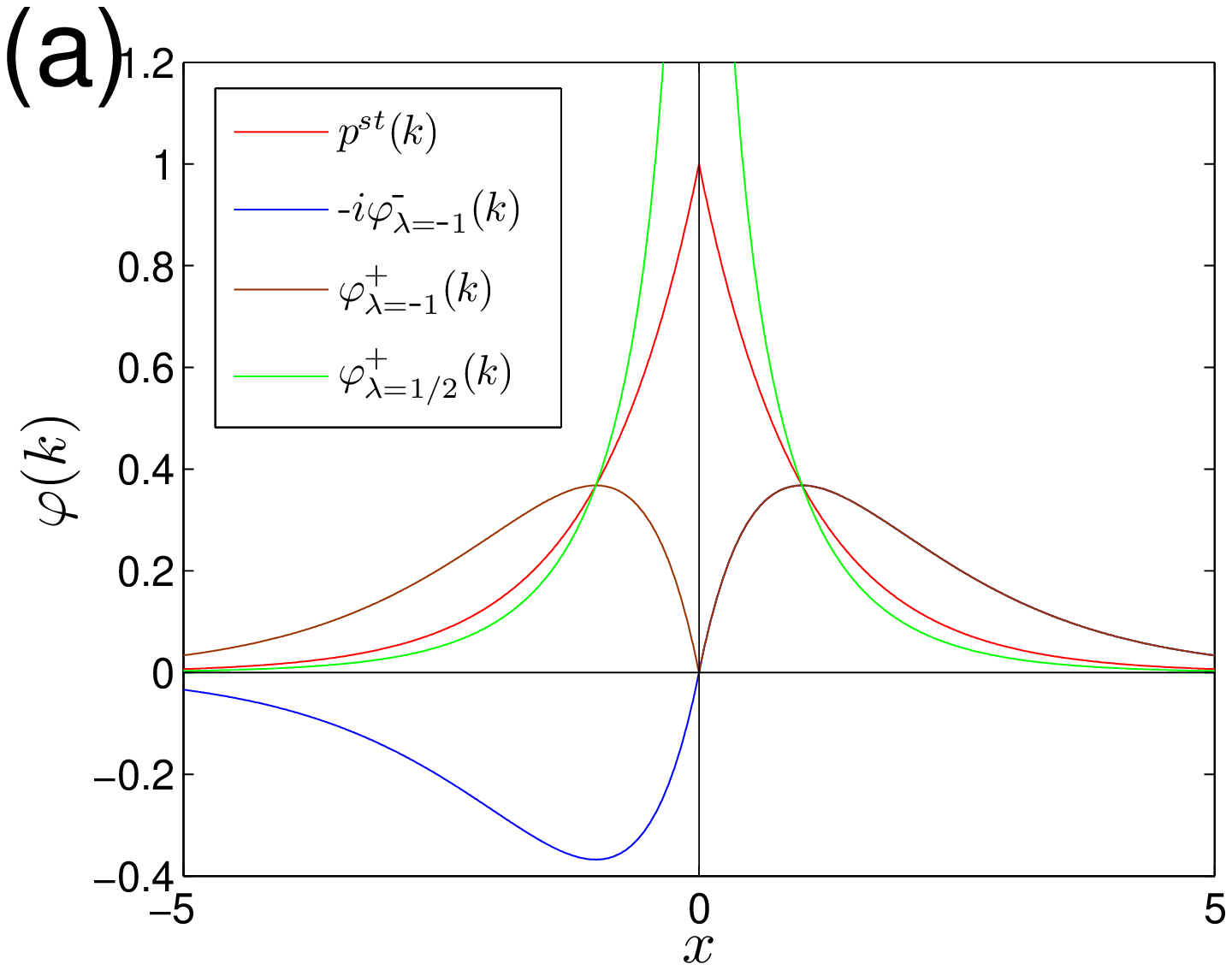}
\includegraphics[width=4.2cm]{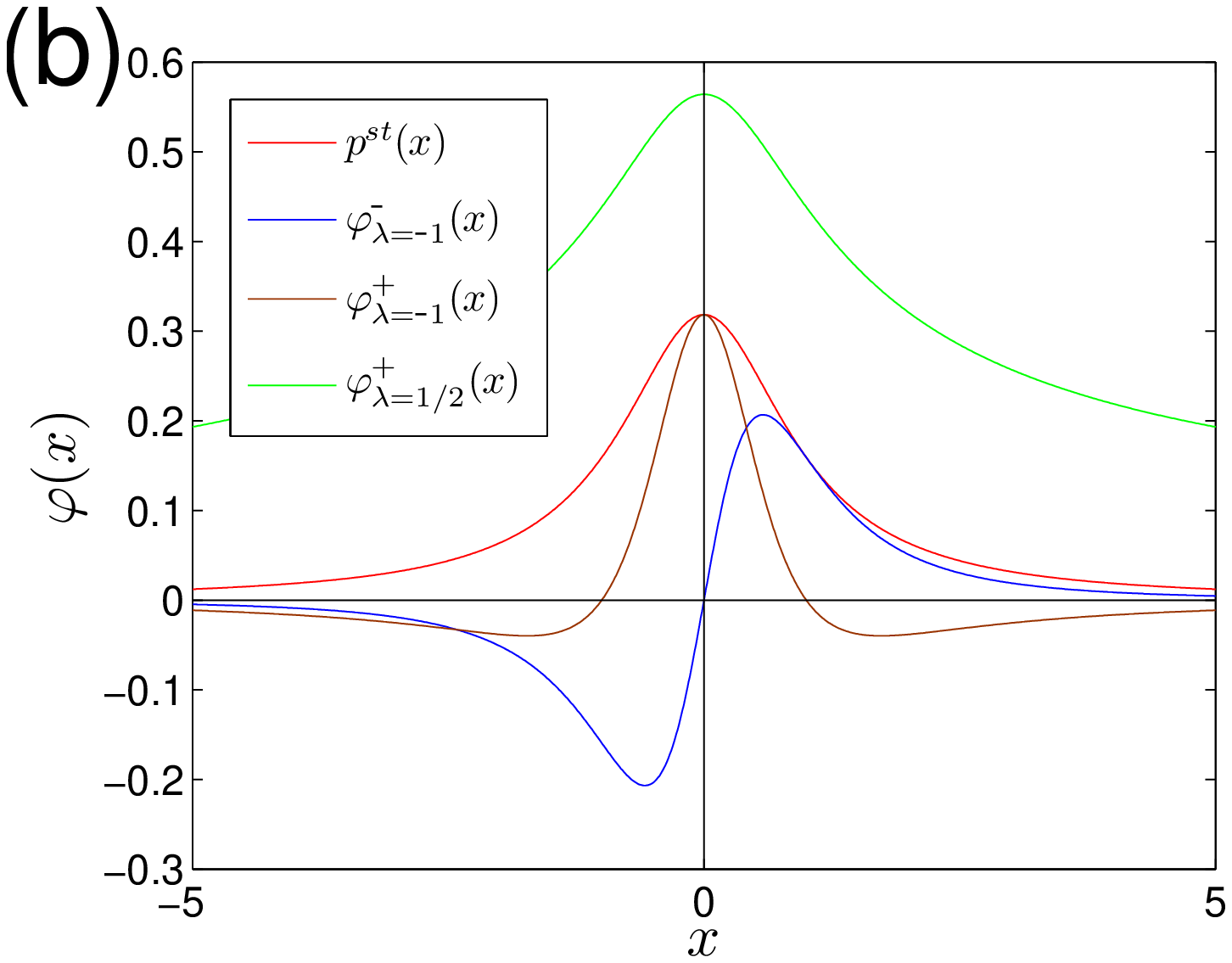}
\caption{Eigenfunctions (\ref{Eq:COUPEigen}) of the Cauchy OUP FP operator with $\mu=1$ (a) in Fourier space and (b) in real space. Shown are the eigenfunctions corresponding to the eigenvalues $\lambda=0$ (even, stationary distribution, red), the odd and even eigenfunctions with $\lambda=-1$ (blue,brown), and the even eigenfunction with $\lambda=1/2$ (green). All eigenfunctions decay asymptotically as powerlaws $\sim x^{\lambda-1}$.}
\label{Fig:COUEigen}
\end{figure}
In real space the transformation kernel (\ref{Eq:DefT}) is obtained as
\begin{eqnarray} \label{Eq:KernelFourier}
	T^\mu_\alpha(\chi,x) 	&=& \frac{1}{2\pi} \int_{-\infty}^\infty e^{-i(\kappa\chi-\alpha^\frac{1}{\mu}\textrm{sign}(\kappa)|\kappa|^\frac{1}{\alpha}x)} d\kappa \nonumber \\ \\
				&=& \frac{1}{\pi}\textrm{Re}\left[\int_0^\infty e^{-i(\kappa\chi-\alpha^\frac{1}{\mu}\kappa^\frac{1}{\alpha}x)} d\kappa\right] \nonumber 
\end{eqnarray}
For $\alpha=1/2$ and $\mu=1$, in the case of the transformation of the FFPO of the Cauchy OUP to the FPO of the Gaussian OUP, 
defining $z=-\textrm{sign}(x) \chi/\sqrt{\pi|x|}$,
\begin{equation} \label{Eq:FresnelG}
	g(z) = \cos\left(\frac{\pi}{2}z^2\right)\left(\frac{1}{2}-C(z)\right)+\sin\left(\frac{\pi}{2}z^2\right)\left(\frac{1}{2}-S(z)\right),
\end{equation}
and the Fresnel integrals
\begin{equation} \label{Eq:FresnelCS}
 	C(x) = \int_0^x \cos\left(\frac{\pi}{2}t^2\right) dt, \quad S(x) = \int_0^x \sin\left(\frac{\pi}{2}t^2\right) dt,
\end{equation}
we obtain the explicit expressions (Appendix)
\begin{equation} \label{Eq:KernelRealSpace_chix}
	T^1_{1/2}(\chi,x) 	= -\textrm{sign}(x)\frac{1}{\chi} zg(z)
\end{equation}
and
\begin{equation} \label{Eq:KernelRealSpace_xchi}
	T^2_2(x,\chi) 	= -\frac{1}{|x|} zg(z).
\end{equation}
The asymptotic behavior of the transformation kernels is found from the scaling of $zg(z)$ \cite{Abramowitz}
\begin{eqnarray} \label{Eq:KernelAsymptotics}
	zg(z) &\sim& O(z), \qquad z\to 0, \\ \nonumber \\
	zg(z) &\sim& O\left(z^{-2}\right), \qquad z\to \infty, \\ \nonumber \\
	zg(z) &\sim& O\left(z\cos\left(\frac{\pi}{2}\left(z^2-\frac{1}{2}\right)\right)\right), ~z\to -\infty.
\end{eqnarray}
\begin{figure}[t!] 
\includegraphics[width=4.3cm]{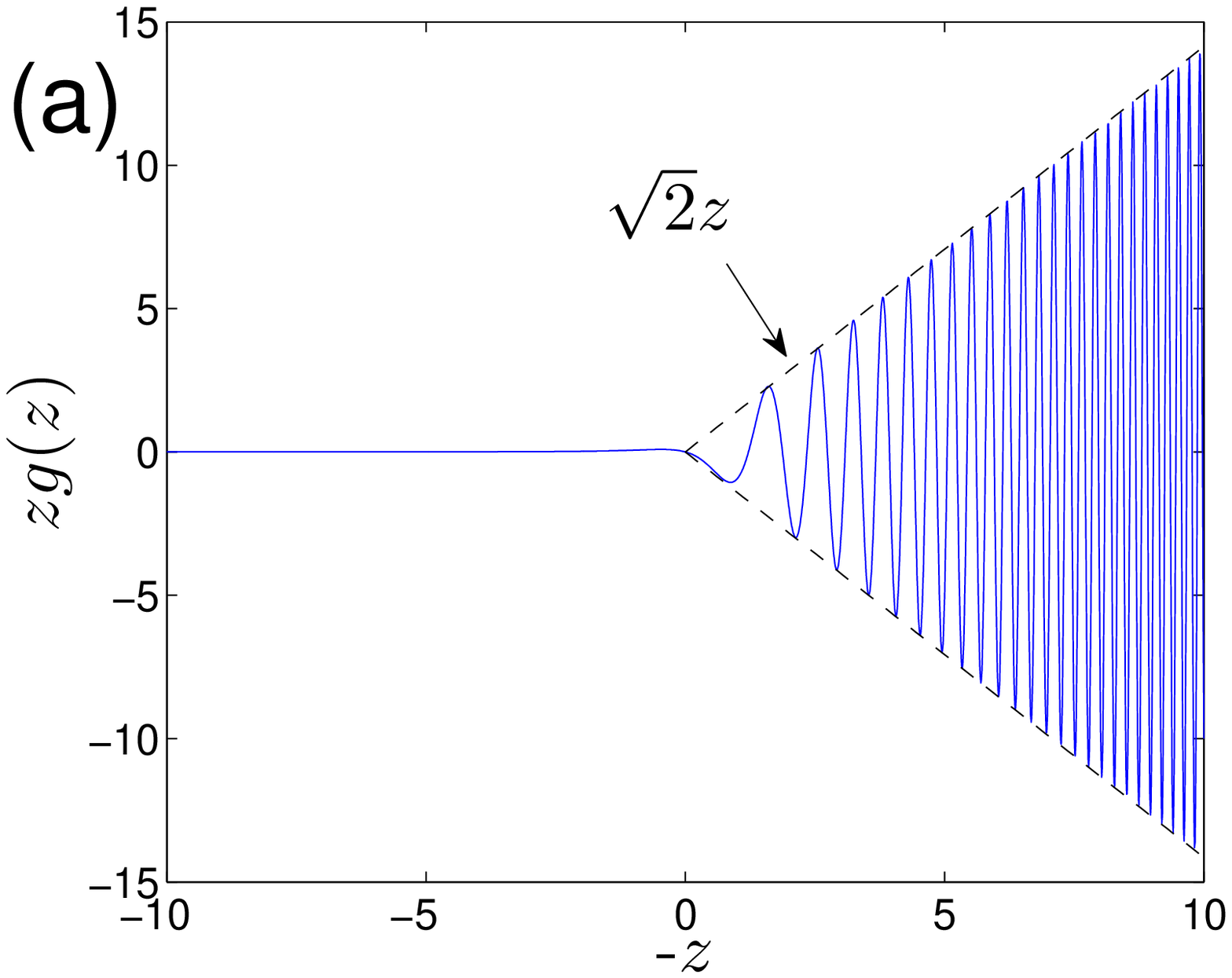}
\includegraphics[width=4.3cm]{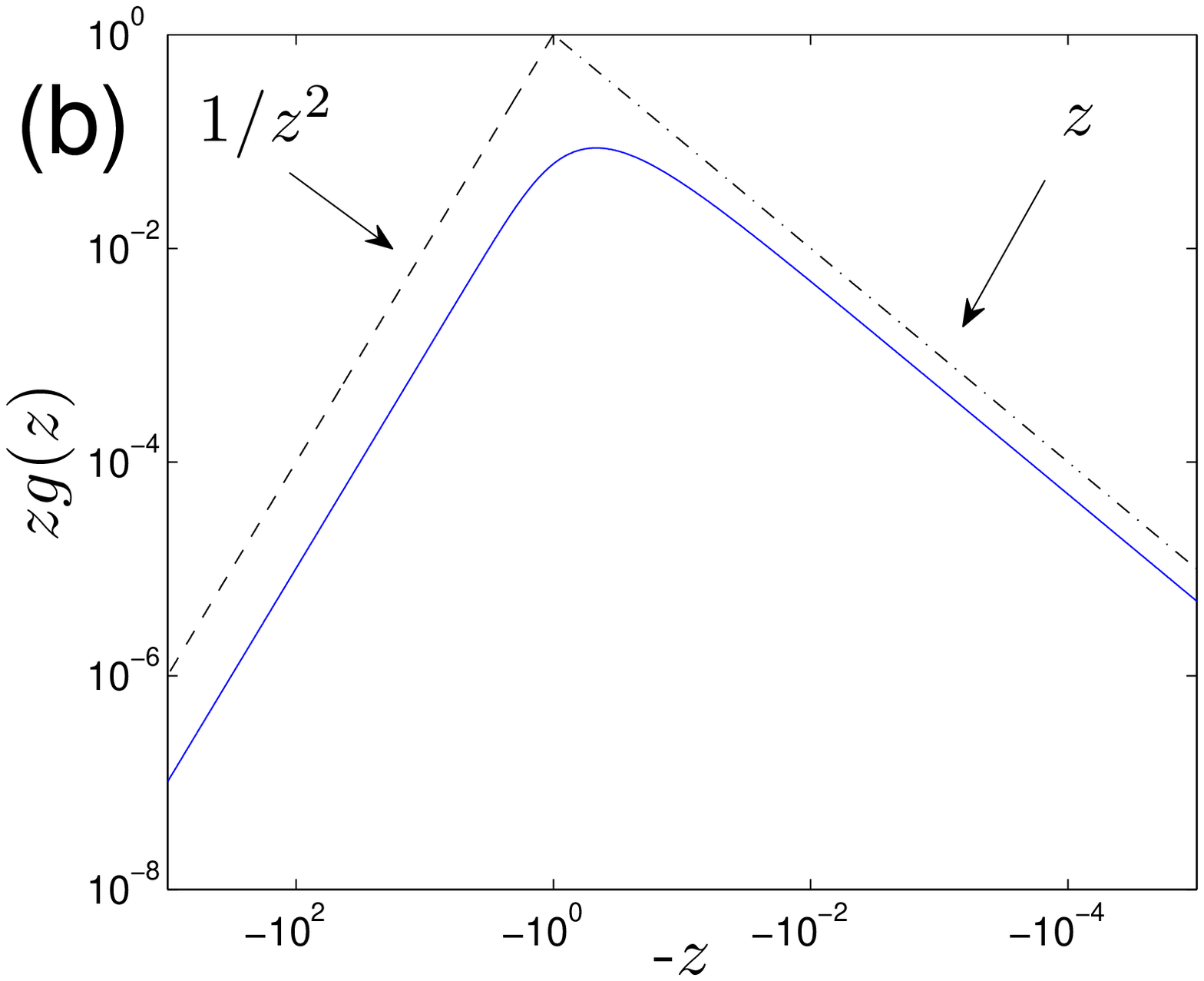}
\caption{The solid blue line shows the scaling function $zg(z)$ of the transformation kernels $T^1_{1/2}(\chi,x)$ and $T^2_2(x,\chi)$ as a function of $-z = \textrm{sign}(x)\chi/\sqrt{\pi|x|}$. Linearly scaled axes in (a) where the dashed line shows the diverging amplitude $\pm\sqrt{2}z$ of the oscillations for $-z\to\infty$. Logarithmically scaled axes in (b) with dashed line indicating the asymptotic scaling with $z^{-2}$ for $z\to\infty$ and the dashed-dotted line the scaling as $z$ for $z\to 0$.}
\label{Fig:zgz}
\end{figure}
\section{Bounded Observables}
The space $L^\infty$ of bounded functions $u(x)$ defines linear functionals $u[f]$ over the space of measurable functions as the Lebesgue-Stieltjes integral $u[f]=\int u(x)f(x)dx$. The functional value $u[\delta(x-x_0)]=u(x_0)$ of a point measure is a random variable $u(t)=u(x_0(t))$ for any realization $x_0(t)$ of a stochastic process. At the same time $u[p]$ is just the finite mean value of $u(x)$ with respect to any probability density function $p(x)$. Since bounded observables project probability densities into the real numbers, relaxation of the component of $p(x,t)$ that is not orthogonal to $u(x)$ can directly be observed. In particular, the families $u^+_{k}(x)=\cos(kx)$ and $u^-_k(x)=\sin(kx)$ constitute a complete set of observables in the sense that its Fourier transform $f(k)=u^+_k[f] + i u^-_k[f]$ determines $f(x)$ almost everywhere. We will therefore study the pointwise relaxation of the characteristic function $p(k,t)$, as well as other bounded observables to the equilibrium. We also look at the relaxation of the correlation functions $c_u(\tau)=\left\langle u(t+\tau)u(t)\right\rangle_t - \left\langle u(t)\right\rangle_t^2$ at equilibrium. 
The difference $\Delta_u(\tau)$ of an observable to its equilibrium value is observed in enseble averages with non-equilibrium initial distributions $p(x,t=0)$, whereas the autocorrelation function $c(\tau)$ of an observable can be measured from a single, stationary realization of a stochastic trajectory. Given a probability density $p(x,t)$ and the conditional probability density $p(x,\tau|x_0)=p(x,t+\tau|x(t)=x_0)$ of a stationary process, the distance of an observable to its equilibrium value and the correlation function are defined by the integrals
\begin{equation}	\label{Eq:DefDeltaX}
	\Delta_u(\tau) 	= \left| \int_{-\infty}^\infty u(x) \left[p(x,\tau)-p^{st}(x)\right] dx \right|
\end{equation}
and
\begin{equation}	\label{Eq:DefCorrX}
	c_{u}(\tau) 	= \int\limits_{-\infty}^\infty  u(x)\left[p(x,\tau|x_0)-p^{st}(x)\right] p^{st}(x_0)u(x_0)	~dx dx_0
\end{equation}
The spectral analysis of (\ref{Eq:DefDeltaX}) and (\ref{Eq:DefCorrX}) assumes that $p(x,\tau)$ and $p(x,t+\tau|x(t)=x_0)$ can be expanded into eigenfunctions of the FFPO and that the integrals can be obtained for each term in the expansion. This is true for $\Delta_u(\tau)$ but not necessarily for $c_u(\tau)$. 
Furthermore, these integrals can be evaluated from Monte-Carlo simulations or observed experimentally. 
Because of finite ensemble sizes and observation time, very large ensembles and very long time series are necessary to observe exponential relaxation to even moderately small logarithmic scales. An ensemble larger than $10^6$ is necessary to resolve differences of an observable from the equilibrium as small as $10^{-3}$. For single trajectory statistics an observation time of $10^6\tau$ is necessary to determine the autocorrelation after time $\tau$ to the same approximate precision. It is therefore possible that only non-spectral relaxation transients may be observed experimentally \cite{Toenjes2013}. A direct numerical evaluation of the integrals with much higher precision is possible by quadrature in Fourier space.
For that, the solution (\ref{Eq:FFPEsolk}) of the FPE for the L\'evy OUP in Fourier space is written as
\begin{figure}[t] 
\includegraphics[width=7cm]{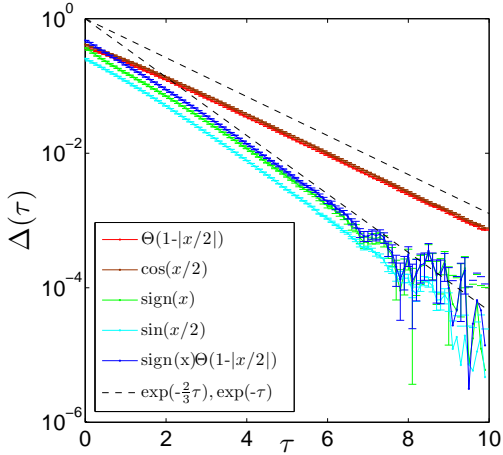}
\caption{Semilogarithmic plots of the difference of observables to their respective equilibrium values in an ensemble of $N=10^8$ independent Gaussian OUPs with $\alpha$-L\'evy-stable distributed ($\alpha=2/3$) initial values with mean shifted to $x_0=1$. Dashed lines are drawn for comparison with the exponential functions $\exp(-\lambda\tau)$ where $\lambda_{001}=-2/3$ and $\lambda_{100}=-1$ are the lowest eigenvalues for even and odd eigenfunctions, respectively. Errorbars indicate ensemble standard deviations.}
\label{Fig:Delta}
\end{figure}
\begin{equation} \label{Eq:FFPESsolutionScaling}
	p(k,t+\tau) = p(ke^{-\tau},t) \frac{p^{st}(k)}{p^{st}(ke^{-\tau})}
\end{equation}
and in particular for the characteristic function of the conditional probability density, with $p(x,t=0)=\delta(x-x_0)$
\begin{equation} \label{Eq:TransitionProbKx0}
	p(k,t+\tau|x(t)=x_0) = e^{ike^{-\tau}x_0} \frac{p^{st}(k)}{p^{st}(ke^{-\tau})}.
\end{equation}
From the properties of convolutions under the Fourier transform follows
\begin{equation}	\label{Eq:distance_general}
	\Delta_u(\tau) 	= \frac{1}{2\pi} \left| \int_{-\infty}^\infty u(-k) p^{st}(k)\left[\frac{p(ke^{-\tau},0)}{p^{st}(ke^{-\tau})}-1\right] dk \right|
\end{equation}
and
\begin{eqnarray}	\label{Eq:correlation}
	c_{u}(\tau) 	&=& \frac{1}{(2\pi)^2} \int\limits_{-\infty}^\infty dkdk' u(-k)u(k') p^{st}(k) \times \\ 
	&&\qquad\times \left[\frac{p^{st}(ke^{-\tau}-k')}{p^{st}(ke^{-\tau})}-p^{st}(-k')\right].	\nonumber 
\end{eqnarray}
\begin{figure}[t] 
\includegraphics[width=7cm]{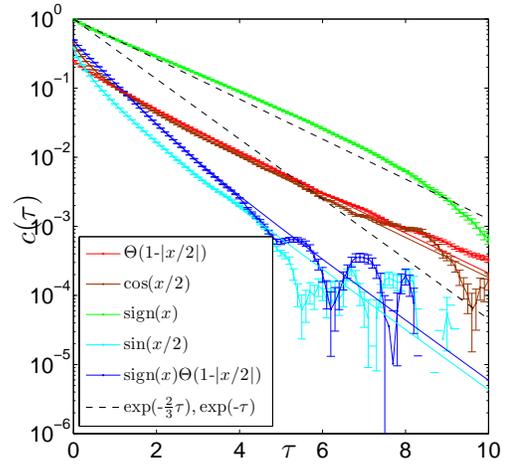}
\caption{Semilogarithmic plots of the autocorrelation function of various even and odd observables for a single trajectory of a L\'evy OUP with $\mu=2/3$ over a dimensionless time of $t=10^6$ with stepsize $\Delta t=0.01$. The solid lines of the same color are evaluated via quadrature in Fourier space (Eq.(\ref{Eq:correlation})). Dashed lines are drawn for comparison with the exponential functions $\exp(-\lambda\tau)$ where $\lambda_{01}=-2/3$ and $\lambda_{10}=-1$ are the lowest eigenvalues for even and odd eigenfunctions, respectively. Error bars indicate moving average standard deviations, but overerestimate the true precision in the presence of exponential (weak) dependence.}
\label{Fig:Corr}
\end{figure}
In contrast to the displacement $\Delta_u$ from equilibrium for $\mu<2$ only bounded observables that decay faster to infinity than any power law show spectral relaxation in their autocorrelation function. This is due to terms proportional to powers of $x_0$ in the expansion of (\ref{Eq:TransitionProbKx0}) into eigenfunctions of the FP operator \cite{Toenjes2013}. Then the average over the initial conditions with respect to the stationary distribution has no spectral decomposition. For instance, the odd observable $\sigma^-(x)=\textrm{sign}(x)$ in the L\'evy OUP with $\mu=2/3$ relaxes to zero at spectral rate $\lambda=-1$ but its autocorrelation does not (see Fig.\ref{Fig:Corr}).

In Fig.\ref{Fig:Delta} we show $\Delta(\tau)$ for the observables $u^+(x)=\cos(x/2)$, $u^-(x)=\sin(x/2)$, $\sigma^-(x)=\textrm{sign}(x)$, $v^+(x)=\Theta(1-|x/2|)$ and $v^-(x)=\sigma^-(x)v^+(x)$ for an ensemble of $N=10^8$ different realizations of Gaussian OUPs with $\mu=2$ and initial values drawn from an $\alpha$-L\'evy-stable distribution with $\alpha=2/3$ and shifted center $x_0=1$. The autocorrelation functions of the same observables calculated from moving avarages along a single realization of a L\'evy OUP with $\mu=2/3$ are shown in Fig.\ref{Fig:Corr}.
The relaxation rates are given by the eigenvalues $\lambda_{lmn}=-(l+ \mu m + \alpha n)$ for observables in the Gaussian OUP ($\mu=2$) with broad initial conditions and by $\lambda_{mn}=-(m+\mu n)$ for the autocorrelation functions of the same observables in the L\'evy OUP ($\mu=2/3$). The observables and their autocorrelation functions decay asymptotically as expected with rates $\lambda_{001}=\lambda_{01}=-2/3$ for the even observables and $\lambda_{100}=\lambda_{10}=-1$ for the odd observables. Note, that in the case of relaxation to the equilibrium from a broad initial distribution in the Gaussian OUP (Fig.\ref{Fig:Delta}) it is the even observable that relaxes at a non-spectral rate $\lambda\ne -n$ with $n\in\mathbb{N}$.  In case of the autocorrelations (Fig.\ref{Fig:Corr}) the odd observables relax at a non-spectral rate $\lambda\ne -n\mu/2$. The autocorrelation function of the bounded observable $\sigma^-(x)=\textrm{sign}(x)$ relaxes at a rate that is not associated with an odd eigenfunction at all.
\section{Conclusions}

Although the Fokker-Planck operator for the Gaussian and the general L\'evy Ornstein-Uhlenbeck process can be mapped by an invertible similarity transformation onto the Hermitian Schr\"odinger operator of the quantum harmonic oscillator, the spectral properties of these operators are vastly different. This is due to the rigid selection rule for the eigenfunctions of the Schrödinger equation which have to be square integrable in real space. The selection rule for the Fokker-Planck operator is much milder. Most admissible, integrable solutions of the Fokker-Planck equation are mapped to diverging, non square integrable functions. The eigenvalue spectra of the Fokker-Planck operators for the classical one dimensional, Gaussian Ornstein-Uhlenbeck process and the fractional Fokker-Planck operator for the L\'evy Ornstein-Uhlenbeck process are trivially continuous with two dimensional eigenspaces spanned by even and odd eigenfunctions. It depends on the problem at hand whether the method of spectral decomposition is applicable and which eigenvalues can be observed in the relaxation of a system to its equilibrium. Further work will be necessary to formalize spectral decomposition of integrable functions into eigenfunctions of the FFPO of the L\'evy OUP without preselecting admissible eigenfunctions, e.g. the ones corresponding to the spectrum of the harmonic oscillator. Our results are directly applicable to the analysis and, most importantly, the interpretation of time series data in non-equilibrium stochastic systems.

\bibliography{EPJB_submit01}

\section{Appendix : Fourier Transforms}

The characteristic function of a probability density function is its Fourier transform defined as
\begin{equation} \label{Eq:DefFourier}
	f(k) = \int_{-\infty}^\infty f(x) e^{ikx} dx.
\end{equation}
Because we assume this convention, the inverse Fourier transform is defined with a prefactor of $1/2\pi$ and negative sign in the exponent as
\begin{equation} \label{Eq:InvFourier}
	f(x) = \frac{1}{2\pi}\int_{-\infty}^\infty f(k) e^{-ikx} dx.
\end{equation}
Multiplication with $-ik$ in Fourier space corresponds to simple differentiation in real space. From that follows that multiplication with $k^{2n}$, $n\in\mathbb{N}$ in Fourier space corresponds to the application of $(-1)^n \partial_x^{2n}$ and multiplication with $ik^{2n+1}$ to the application of $(-1)^{n+1} \partial_x^{2n+1}$ in real space. For the special convolution that is the inner product we have in our convention
\begin{equation} \label{Eq:Convolution}
	\int_{-\infty}^\infty f(x)g(x) dx = \frac{1}{2\pi} \int_{-\infty}^\infty f(-k)g(k) dk.
\end{equation}
The characteristic function $p^{st}(k=0)$ of any probability density function should be equal to unity. Other than this, we did not consider any orthonormalization of eigenfunctions, so the prefactors of the eigenfunctions are arbitrary. In this paper the eigenfunctions of the FFPO were specified as
\begin{equation} \label{Eq:EvenEigenfuncAppendix}
	\varphi_\lambda^+ = |k|^{-\lambda}e^{-\frac{1}{\mu}|k|^\mu}
\end{equation}
and
\begin{equation} \label{Eq:OddEigenfuncAppendix}
	\varphi_\lambda^- = i\textrm{sign}(k)|k|^{-\lambda}e^{-\frac{1}{\mu}|k|^\mu}.
\end{equation}
The inverse Fourier transforms of these eigenfunction in the Cauchy case $\mu=1$ are given exactly in Eq. (\ref{Eq:COUPEigen}). In the Gaussian case with $\mu=2$ the inverse Fourier transforms under the convention (\ref{Eq:InvFourier}) and drawn in Fig. \ref{Fig:GaussEigen} are
\begin{eqnarray} \label{Eq:GaussFPOEigenfuncAppendix}
	\varphi_\lambda^+(x) &=&  M\left(\frac{1}{2}\left[1-\lambda\right];\frac{1}{2}; -\frac{1}{2}x^2\right)\frac{1}{2\pi}2^{\frac{1}{2}(1-\lambda)}\Gamma\left(\frac{1}{2}(1-\lambda)\right)  \nonumber \\ \\
	\varphi_\lambda^-(x) &=&   x M\left(1-\frac{\lambda}{2};\frac{3}{2}; -\frac{1}{2}x^2\right)\frac{1}{2\pi}2^{1-\frac{\lambda}{2}}\Gamma\left(1-\frac{\lambda}{2}\right). \nonumber
\end{eqnarray}
The transformation kernels $T^1_{1/2}(\chi,x)$  and $T^2_2(x,\chi)$ in real space are obtained from the inverse Fourier transform as
\begin{eqnarray} 
	T^1_{\frac{1}{2}}(\chi,x) 	&=& \frac{1}{2\pi} \int_{-\infty}^\infty e^{-i(\kappa\chi-\frac{1}{2}\textrm{sign}(\kappa)|\kappa|^2x)} d\kappa \nonumber \\ \\
				&=& \frac{1}{\pi}\textrm{Re}\left[\int_0^\infty e^{-i(\kappa\chi-\frac{1}{2}\kappa^2x)} d\kappa\right] \nonumber 
\end{eqnarray}
Substituting 
\begin{equation}\label{Eq:SubChiApp}
	\xi=\sqrt{\frac{2}{\pi}}\left(\sqrt{\frac{|x|}{2}}\kappa - \textrm{sign}(x)\frac{\chi}{\sqrt{2|x|}}\right)
\end{equation}
and 
\begin{equation} \label{Eq:SubzApp}
	z=- \textrm{sign}(x)\frac{\chi}{\sqrt{\pi|x|}}
\end{equation}
we have 
\begin{equation}
	d\kappa = \sqrt{\frac{\pi}{|x|}}d\xi, \qquad \xi(\kappa=0)=z
\end{equation}
and
\begin{equation}
	\frac{1}{2}\kappa^2x-\kappa\chi = \textrm{sign}(x)\left(\frac{\pi}{2}\xi^2-\frac{\chi^2}{2|x|}\right)
\end{equation}
so that
\begin{eqnarray} 
	T^1_{\frac{1}{2}}(\chi,x) 	&=& 
			\frac{1}{\sqrt{\pi|x|}} \textrm{Re}\left[ e^{-i\frac{\chi^2}{2x}} \int_z^\infty e^{i\textrm{sign}(x)\frac{\pi}{2}\xi^2} d\xi\right].
\end{eqnarray}
With $\frac{1}{\sqrt{\pi|x|}}=-\textrm{sign}(x)\frac{z}{\chi}$, $-\frac{\chi^2}{2x}=-\textrm{sign}(x)\frac{\pi}{2}z^2$, the Fresnel integrals
\begin{equation} \label{Eq:FresnelCSAppendix}
 	C(x) = \int_0^x \cos\left(\frac{\pi}{2}t^2\right) dt, \quad S(x) = \int_0^x \sin\left(\frac{\pi}{2}t^2\right) dt,
\end{equation}
and
\begin{equation}
	\int_z^\infty e^{i\textrm{sign}(x)\frac{\pi}{2}\xi^2}d\xi = \left(\frac{1}{2}-C(z)\right) +i\textrm{sign}(x)\left(\frac{1}{2}-S(z)\right)
\end{equation}
the transformation kernel is found to be
\begin{equation}
	T^1_{\frac{1}{2}} = -\textrm{sign}(x)\frac{1}{\chi}zg(z)
\end{equation}
where
\begin{equation} \label{Eq:FresnelGAppendix}
	g(z) = \cos\left(\frac{\pi}{2}z^2\right)\left(\frac{1}{2}-C(z)\right)+\sin\left(\frac{\pi}{2}z^2\right)\left(\frac{1}{2}-S(z)\right).
\end{equation}
The inverse transform $T^2_2$ is calculated as
\begin{eqnarray} 
	T^2_{2}(x,\chi) 	&=& \frac{1}{2\pi} \int_{-\infty}^\infty e^{-i(k x-\textrm{sign}(k)\sqrt{2|k|}\chi)} dk \nonumber \\ \\
				&=& \frac{1}{\pi}\textrm{Re}\left[\int_0^\infty e^{-i(kx-\sqrt{2k}\chi)} dk\right] \nonumber 
\end{eqnarray}
Substituting $k=\frac{1}{2}\kappa^2$, i.e. $dk = \kappa d\kappa$ we find
\begin{eqnarray} 
	T^2_{2}(x,\chi) 	&=& \frac{1}{\pi}\textrm{Re}\left[\int_0^\infty \kappa e^{-i(\frac{1}{2}\kappa^2 x-\kappa \chi)} d\kappa\right] \nonumber \\ \nonumber \\
			&=& \frac{1}{\pi}\textrm{Re}\left[\int_0^\infty \kappa e^{-i(\kappa \chi-\frac{1}{2}\kappa^2 x)} d\kappa\right] \nonumber \\ \nonumber \\
			&=& \frac{1}{\pi}\textrm{Re}\left[i \partial_\chi \int_0^\infty  e^{-i(\kappa \chi-\frac{1}{2}\kappa^2 x)} d\kappa\right]. 
\end{eqnarray}
Making again the substitutions (\ref{Eq:SubChiApp},\ref{Eq:SubzApp}) we have
\begin{eqnarray}
	T^2_2(x,\chi) &=& \frac{1}{\sqrt{\pi|x|}}\textrm{Re}\left[i \partial_\chi z \partial_z \left(e^{-i\textrm{sign(x)}\frac{\pi}{2}z^2} \int_z^\infty  e^{i\textrm{sign}(x)\frac{\pi}{2}\xi^2}d\xi\right)\right]. \nonumber \\ \nonumber \\
			&=& \frac{\chi}{x} T^1_{1/2}(\chi,x) = -\frac{1}{|x|}zg(z).
\end{eqnarray}

\end{document}